\documentclass[reprint,nofootinbib,amsmath,amssymb,aps]{revtex4-2}

\usepackage{graphicx} 

\usepackage{amssymb}
\usepackage{amsmath}
\usepackage{subfigure}
\usepackage{enumitem}
\usepackage{color}
\usepackage{hyperref}
\usepackage{float}

\newcommand{\hMpc}{$ \, h^{-1}  \rm Mpc$}

\newcommand{\hkpc}{$ \, h^{-1} \rm kpc$}
\newcommand{\hpc}{$ \, h^{-1} \rm pc$}

\newcommand{\hMsun}{$\, h^{-1} \rm M_{\odot}$}

\newcommand{\mvir}{$M_{\rm vir}$}
\newcommand{\rvir}{$R_{\rm vir}$}

\newcommand\apjl{{\it Astrophys. J. Lett.}}

\newcommand\aap{{\it Astron. Astrophys.}}
\newcommand\mnras{{\it Mon. Not. R. Astron. Soc.}}
\newcommand\pasj{{\it Publ. of the Astron. Society of Japan}}
\newcommand\na{{\it New Astronomy}}
\newcommand\jcap{J. Cosmology Astropart. Phys}

\newcommand{\sidmi}{$\rm\ SIDM_1$}
\newcommand{\sidmii}{$\rm\ SIDM_3$}
\newcommand{\rhoc}{$\rho (\rm 150 pc)$}
\newcommand{\vpeak}{$V_{\rm peak}$}
\newcommand{\vmax}{$V_{\rm max}$}
\newcommand{\rperi}{$r_{\rm peri}$}

\newcommand{\ufdpaper}{Hayashi~et.~al. (in prep.)}

\begin{document}
\title{Constraining Self-Interacting Dark Matter with Dwarf Spheroidal Galaxies and High-resolution Cosmological $N$-body Simulations}

\author{Toshihiro Ebisu}
\affiliation{Department of Applied and Cognitive Informatics, Chiba University, 1-33, Yayoi-cho, Inage-ku, Chiba 263-8522, Japan}
\author{Tomoaki Ishiyama}
\email[e-mail:]{ishiyama@chiba-u.jp}
\affiliation{Institute of Management and Technologies, Chiba University, 1-33, Yayoi-cho, Inage-ku, Chiba 263-8522, Japan}
\author{Kohei Hayashi}
\affiliation{National Institute of Technology, Ichinoseki College, Takanashi, Hagisho, Ichinoseki, Iwate, 021-8511, Japan}

\preprint{}
\date{\today}


\begin{abstract}
We study the density structures of dark matter subhalos for both cold
dark matter and self-interacting dark matter models using
high-resolution cosmological $N$-body simulations.  
We quantify subhalo's central density at 150 pc from the center of each subhalo
at the classical dwarf spheroidal and ultrafaint dwarf scales 
found in Milky-Way sized halos.
By comparing them with observations, we find that the
self-interacting scattering cross-section of
$\sigma/m<3\ \rm{cm^{2}g^{-1}}$ is favored.  
Due to the combination of hosts' tide and self-interactions,
the central density of
subhalos with small pericenter shows a noticeable difference
between the cold and the self-interacting models, indicating that
Milky-Way satellites with small pericenter are ideal sites to
further constrain the nature of dark matter by future large
spectroscopic surveys.
\end{abstract}


\maketitle

\section{Introduction}\label{sec_introduction}
The cold dark matter (CDM) model, which assumes collisionless dark matter
particles, has successfully reproduced the structure of the Universe
on large scales ($\geq 1\ \rm{Mpc}$)~\cite{Bahcall1999, Tegmark2004, Springel2006, Mandelbaum2018, Heymans2021, Ishiyama2021, DES2021}.
However, on smaller scales, especially on
dwarf galaxies, discrepancies exist between observations and CDM
predictions, called the ``small scale crisis''.  For example,
cosmological $N$-body simulations based on the CDM predict cuspy
density profiles of dark matter halos~\cite{Navarro1997}, 
whereas observations
of dwarf galaxies suggest cored profiles~\cite{burkert1995structure,
  de2003simulating, oh2011dark, oh2011central}.  This discrepancy is
known as the core-cusp problem and has recently been re-interpreted as the
diversity problem~\cite{Oman2015}: circular velocity profiles
of observed dwarf galaxies show a large diversity
even with similar maximum circular velocity. 
Some dwarf galaxies have cuspy profiles, and others have cored
profiles, whereas those simulated galaxies show little variation. 
Baryonic physics in galaxies can be a solution to this problem
although their effects are still uncertain. 

Self-interacting dark matter (SIDM) models~\cite{Spergel2000} have
been proposed to solve the small scale crisis (for a review, see
\cite{tulin2018dark}).  The original SIDM assumes that the scattering
cross-section of dark matter particles per unit mass, $\sigma/m$, is
velocity-independent and isotropic.  However, preferred cross-sections
are different depending on scales: $\sigma/m \gtrsim 1 \, \rm
cm^{2}g^{-1}$ for dwarf 
galaxies~\cite{Zavala2013, Elbert2015, kaplinghat2016dark} and 
$\sigma/m \lesssim 0.1 \mbox{\scriptsize --} 1 \, \rm cm^{2}g^{-1}$ 
for galaxy clusters~\cite{kaplinghat2016dark, kim2017wake, robertson2017what, sagunski2021velocity}.
To overcome
this hurdle, reincarnations of the original SIDM have been considered,
such as velocity-dependent SIDM 
(e.g.,~\cite{loeb2011cores, Vogelsberger2012, Chu:2018fzy, Banerjee2020, Nadler2020}) and SIDM with
anisotropic scattering from the approach of particle physics
(e.g.,~\cite{ackerman2009dark, Petraki2013, kahlhoefer2014colliding}).

Dark matter self-interaction decreases the central density of halos 
before gravothermal core-collapse sets in~\citep{Balberg:2002ue,Nishikawa:2019lsc}. 
Baryonic physics in galaxies, such as supernova feedback and 
star formation burst, could also affect the central structures of halos. 
These two different mechanisms can coincide and interplay. 
To remove this degeneracy, 
it is ideal for studying dark matter dominated systems such as 
classical dwarf spheroidal (dSph) and ultrafaint dwarf (UFD) galaxies. 
Cosmological hydrodynamical simulations predict that 
central structures of those host halos are less affected by feedback processes~\cite{Tollet2016, Fitts2017, Lazar2020}, 
maybe because most of the stars in those system were born before the cosmic reionization.

Only recently, the density structures of UFDs have been
estimated, and the SIDM cross-section is constrained by them~\cite{Hayashi2021}.  However, a phenomenological 
semi-analytic SIDM halo model is used
in that study because there is a significant lack of high-resolution
simulations for UFD scale subhalos. 
Semi-analytic models have been 
used to predict density structures of isolated halos 
using various SIDM 
models~\cite{Kaplinghat2016, Kamada2017, Valli2018, Hayashi2021, Robertson2021}, 
and have the advantage of including baryonic
potential and being less computational cost than cosmological simulations.
However, in semi-analytic models, it is challenging to model 
the dynamical evolution of subhalos within a host halo, 
such as tidal interactions with host halos and other subhalos, 
and evolution of host halos. 
Cosmological simulations have the advantage that all these dynamical 
effects are naturally included. 

This paper will revisit the SIDM cross-section using high-resolution
cosmological $N$-body simulations based on both CDM and SIDM, and
observations of the classical dSphs and the UFDs.  Unlike previous
studies that focused on the diversity of circular velocity profiles
(e.g.,~\cite{Zavala2019}), we focus on density structures of the
classical dSph and the UFD scales subhalos.  We quantify the central
density at 150 pc, \rhoc, from the center of each subhalo and compare
them with observations. \citet{read2019dark} argued that \rhoc\ is a
good tracer of the central density of dwarf galaxies because it is
insensitive to the prior choice of the inner slope of the density
profile and shows enough difference to distinguish whether the density
profile is cusp or cored.  Uncertainty in \rhoc\ is not prohibitively
large, 
and 1$\sigma$ uncertainty is typically less than 50\% 
for the classical dSphs~\cite{read2019dark}. 
The major source of uncertainty is poor observed sample size, thus, 
future spectroscopic surveys can improve the uncertainty.
We also study the dependence of central density on the
pericenter distance; It has been pointed out that the dSph's central
density \rhoc\ anti-correlates with the pericenter radii
\cite{Kaplinghat2019, Hayashi2020}.

This paper is organized as follows.  In Section~\ref{sec_simulations},
we explain the details of our cosmological $N$-body simulations.  In
Section~\ref{sec_results} and \ref{sec_discussion}, we show and discuss the results. 
Finally, we present the conclusions in Section~\ref{sec_conclusions}.

\section{Simulations}\label{sec_simulations}

We performed three high-resolution cosmological $N$-body simulations:
CDM (without self-interaction), \sidmi\ ($\sigma/m=1\ \rm{cm^{2} \,
  g^{-1}}$), and \sidmii\ ($\sigma/m=3 \, \rm{cm^{2}\ g^{-1}}$).  All
of the simulations consist of $1024^{3}$ dark matter particles with
a particle mass of $4.1\times10^{4}$\hMsun\ in a comoving cubic box
with a side length of 8\hMpc.  We constructed an initial condition
using \textsc{2LPTic} code~\cite{Crocce2006}, and the adopted
cosmological parameters are $\Omega_{0}=0.31$, $\Omega_{\rm b}=0.048$,
$\lambda_{0}=0.69$, $h=0.68$, $n_{\rm s}=0.96$, and $\sigma_{8}=0.83$,
which match with observational results by the Planck~\cite{Planck2020}.  We used the same initial condition and numerical
parameters for three simulations, in which only the strength of
self-interactions is different.

To follow the gravitational evolution of dark matter particles, we
used a massively parallel TreePM code,
\textsc{GreeM}~\footnote{\url{https://hpc.imit.chiba-u.jp/~ishiymtm/greem/}}
\cite{Ishiyama2009b,Ishiyama2012}, on the Aterui-II supercomputer at the Center for
Computational Astrophysics, National Astronomical Observatory of
Japan.  The gravitational softening length is 100\hpc.  We accelerated
the tree force calculation by the
\textsc{Phantom-grape}\footnote{\url{https://bitbucket.org/kohji/phantom-grape/src}}
software~\citep{Nitadori2006, Tanikawa2012, Tanikawa2013,
  Yoshikawa2018}.  We performed all simulations from redshift $z=127$
to 0 and stored snapshots covering $z=12$ to 0 with a logarithmic
interval $\Delta \log(1+z)=0.01$, resulting in a total of 112
snapshots. We adopted the same algorithm used in~\citet{Vogelsberger2012}
for the implementation of dark matter self-interaction, in which
isotropic, velocity independent, and elastic scattering are
considered.  We do not consider 
velocity-dependent SIDM models because we focus on scales smaller than galaxies. 
In Appendix, we compare the core size of the host halos with a
previous study to show the validity of our SIDM implementation.

We identified gravitationally bound dark matter halos and subhalos using \textsc{rockstar} halo/subhalo finder
\cite{Behroozi2013} and constructed merger trees using
\textsc{consistent trees} merger tree code~\cite{Behroozi2013b}. 
We unbiasedly picked out nine Milky Way-sized host halos (H0-H8) covering the
virial mass range of $3.4\times10^{11}$ to
$2.8\times10^{12}$\hMsun. The virial mass \mvir\ and radius
\rvir\ of each halo are summarized in Table~\ref{tab:halos}.  
We analyzed their subhalos more massive than $1.0 \times 10^{8}$\hMsun, 
which consists of more than 2,400 particles 
and corresponds to the massive UFD scale.

\begin{table*}
\begin{center}
\small
\caption{Virial mass \mvir\ and radius \rvir\ of nine Milky Way-sized host halos.}
\label{tab:halos}
\begin{tabular}{c | c c | c c | c c} \hline
      & CDM &  & $\rm SIDM_{1}$ &  & $\rm SIDM_{3}$ & \\ \hline
 Name & $M_{\rm{vir}}$ & $R_{\rm{vir}}$ & $M_{\rm{vir}}$ & $R_{\rm{vir}}$ & $M_{\rm{vir}}$ & $R_{\rm{vir}}$\\ 
      & [$10^{11}$\hMsun] & [\hkpc] & [$10^{11}$\hMsun] & [\hkpc] & [$10^{11}$\hMsun] & [\hkpc]\\ \hline
 H0 & 25.76  & 278.6 & 25.65 & 278.2 & 24.94 & 275.6\\
 H1 & 11.08  & 210.3 & 11.04 & 210.0 & 10.97 & 209.6\\
 H2 & 10.56  & 206.9 & 10.51 & 206.6 & 10.37 & 205.7\\
 H3 & 8.105 & 189.4 & 7.920 & 188.0 & 7.687 & 186.2\\
 H4 & 7.141 & 181.7 & 7.097 & 181.2 & 7.033 & 180.7\\
 H5 & 6.902 & 179.6 & 6.861 & 179.3 & 6.795 & 178.7\\
 H6 & 6.800 & 178.7 & 6.873 & 179.4 & 6.732 & 178.1\\
 H7 & 4.581 & 156.7 & 4.520 & 156.0 & 4.435 & 155.0\\
 H8 & 3.755 & 146.6 & 3.556 & 144.0 & 3.463 & 142.7\\\hline
\end{tabular}
\end{center}	
\end{table*}

\section{Results}\label{sec_results}

\subsection{Central density of subhalos}\label{sec_rho150}

\begin{figure*}
\includegraphics[width=14cm]{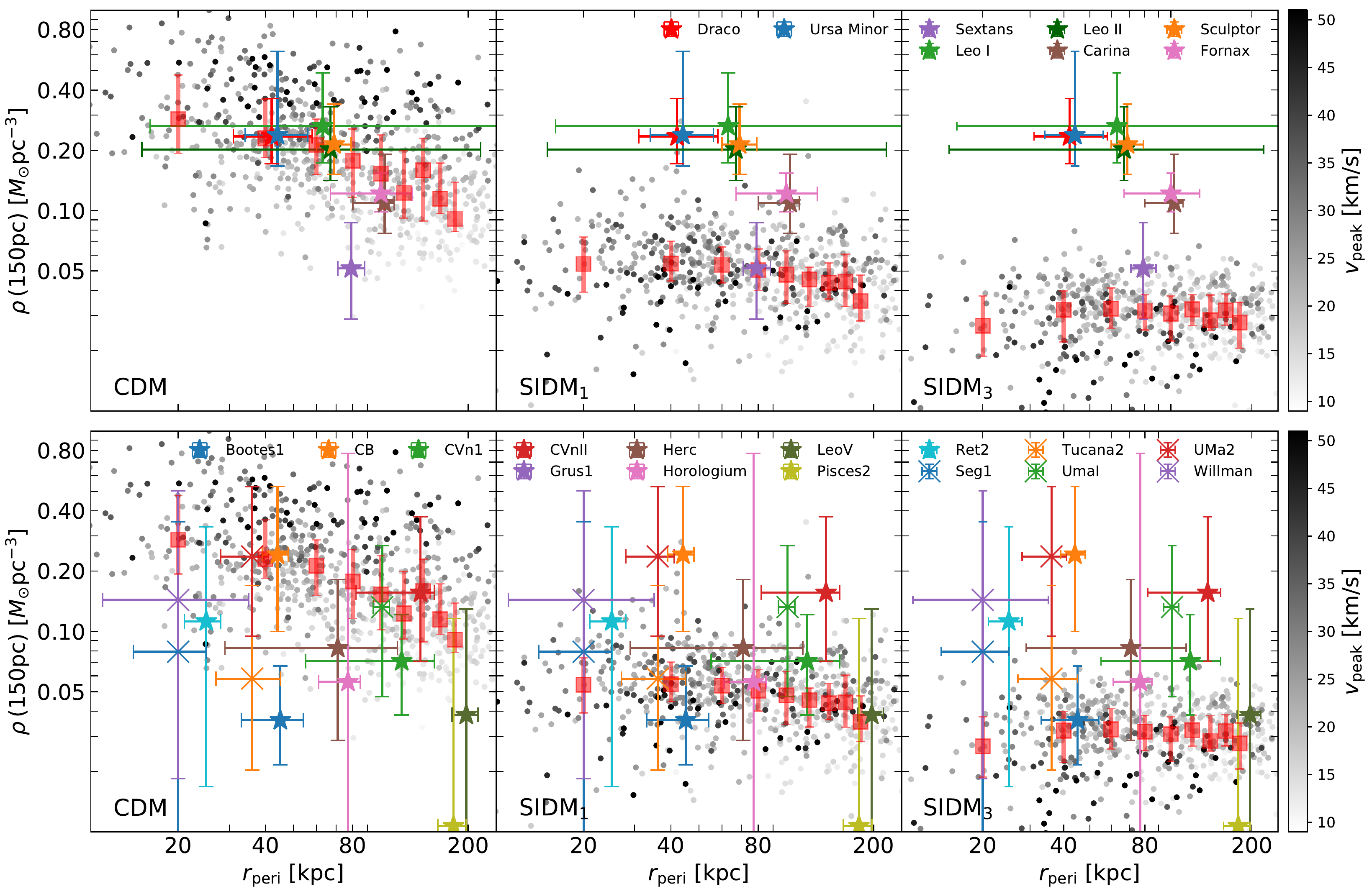}
\caption{
Subhalos' density at 150 pc, \rhoc, versus their
pericenter radii, \rperi. 
Color filled stars and crosses with error
bars show observational results of the eight classical dSphs
\cite{Hayashi2020} (Draco, Ursa Minor, Sextans, Leo I, Leo II, Sculptor, Carina, and Fornax are shown in the upper row) 
and the 15 UFDs (Bo\"{o}tes, Coma Berenices (CB), Canes Venatici I (CvnI), Canes Venatici II (CvnII), Hercules (Herc), Leo V, Grus 1, Horologium I, Pisces II, Reticulum II (Ret2), Tucana 2, Ursa Major I (UMa1), Ursa Major II (UMa2), Segue 1 (Seg1), and Willman 1 
are shown in the bottom row) (\ufdpaper).
Filled circles are individual subhalos in the nine Milky Way-sized
halos in simulations, CDM (left), \sidmi\ (middle), and
\sidmii\ (right).  The densities at 150 pc are estimated by fitting with
the NFW profile for the CDM simulation and the Burkert profile for the
\sidmi\ and \sidmii\ simulations.  The contrast of squares is proportional to the
\vpeak, which is the maximum circular velocity over the subhalo's
history.  Red filled squares with error bars represent the first and
third quartiles of \rhoc\ in each pericenter bin.  }
\label{fig:rho150_1}
\end{figure*}

Fig.~\ref{fig:rho150_1} shows the relation between the subhalos'
density at 150 pc, \rhoc,  and their pericenter radii
$r_{\rm{peri}}$. 
We can not directly measure \rhoc\ from our simulations
because the softening length adopted is 100 \hpc~($=147$ pc). 
Therefore, we estimate \rhoc\ 
by fitting the density profiles of the subhalos with the NFW profile~\cite{Navarro1997} expressed by
\begin{eqnarray}
\rho_{\rm NFW}(r) &=& \frac{\rho_{\rm s}r_{\rm s}^3}{r(r_{\rm s}+r)^2},
\label{eq:nfwprf}
\end{eqnarray}
and with the Burkert profile~\cite{burkert1995structure} expressed by
\begin{eqnarray}
\rho_{\rm{B}}(r) &=& \frac{\rho_{\rm b}r_{\rm b}^3}{(r+r_{\rm b})(r^2+r_{\rm b}^2)}. 
\label{eq:burkertprf}
\end{eqnarray}
We use the NFW profile for the CDM simulation 
and the Burkert profile for the \sidmi\ and \sidmii\ simulations. 
The fittings were performed using up to 20 bins covering 
$0.01R_{\rm vir} \leq r \leq$ \rvir, 
and only bins with $r > 300 \, \rm pc$ were used.
We also calculated the pericenter of each
subhalo as the minimum distance from its host over the subhalo's history. 
The nine pericenter bins cover $10\ \rm{kpc} \leq r_{\rm{peri}} < 190\ \rm{kpc}$ 
at equally spaced intervals. 

We compare simulated \rhoc\ with the observation. We use \rhoc\ of 
observed classical dSphs and UFDs that were estimated by~\cite{Hayashi2020} and \ufdpaper, respectively. 
To compute \rhoc, they determined a dark matter density profile by non-spherical dynamical models based on axisymmetric Jeans equations.
They applied these models to the line-of-sight velocity data for the dSphs and UFDs and obtained the posterior distribution functions for the dark matter halo parameters (see~\cite{Hayashi2020} in details).
Thus, the estimated \rhoc\ were marginalized by all these parameters.

Fig.~\ref{fig:rho150_1} (top row) shows that the central density
\rhoc\ of subhalos in the CDM simulation is in good agreement with
those of the classical dSphs and anti-correlates with the pericenter
\rperi, consistent with observational results~\cite{Kaplinghat2019,
  Hayashi2020}.  On the other hand, the density \rhoc\ of subhalos in
the SIDM simulations is systematically lower than CDM counterparts 
and the classical dSphs.  The central density does not depend
strongly on the pericenter and is lower in the \sidmii\ than in the
\sidmi.  The difference with the classical dSphs is also larger in the
\sidmii.  These results indicate that the SIDM scattering cross
section of $\sigma/m<1\ \rm{cm^{2}g^{-1}}$ is favored.

The bottom row of Fig.~\ref{fig:rho150_1} shows the comparison with
the UFDs.  Unlike the classical dSphs, the central density
\rhoc\ estimated in the observed UFDs does not clearly depend
on the pericenter, although observational uncertainty is large.  Within
the first and third quartiles, none of the three simulations is in good
agreement with the observations.  In terms of the overall
distribution, \sidmi\ shows the best match with the observations.  The
CDM simulation has difficulty reproducing Bo\"{o}tes 1, although it shows
consistency with the others.  For a few UFDs, the \rhoc\ in
\sidmii\ is too low compared to the UFDs.  These results indicate that
the SIDM scattering cross-section of $\sigma/m\lesssim
3\ \rm{cm^{2}g^{-1}}$ is favored, which is consistent with the
indication from the comparison with the classical dSphs.

Fig.~\ref{fig:rho150_1} also shows that 
the density \rhoc\ of subhalos in simulations depend on their \vpeak, 
which is the maximum circular velocity over the subhalo's history. 
Denser subhalos tend to have lower \vpeak\, 
in the \sidmi\ and \sidmii\ simulations,
whereas denser subhalos
tend to have higher \vpeak\ in the CDM simulation.
We further investigate these results in the next section. 

\begin{figure*}
 \includegraphics[width=14cm]{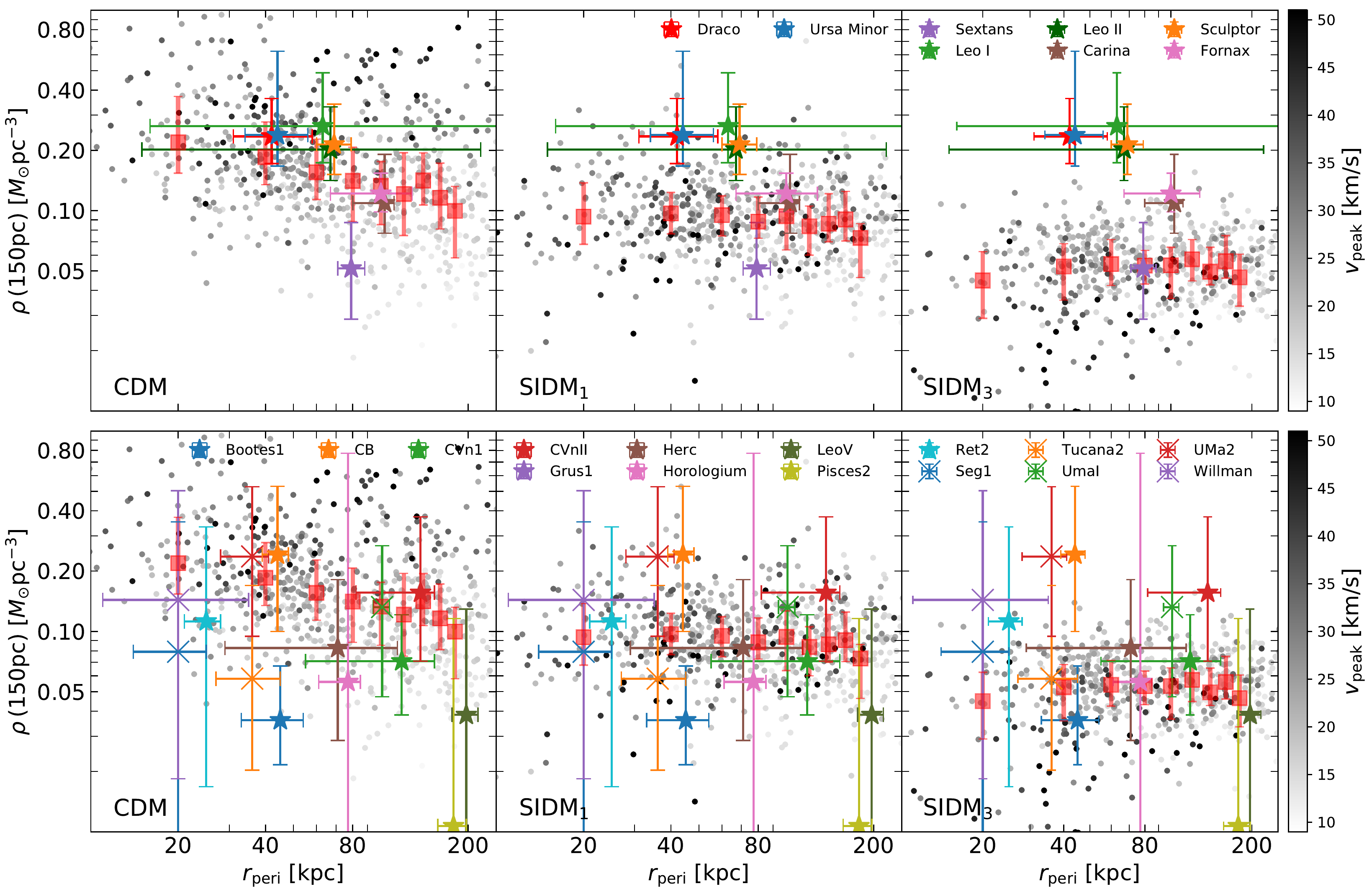}
\caption{Same as Fig.~\ref{fig:rho150_1}, but the densities at 150 pc are 
estimated by fitting with the cuspy profile for all simulations.}
\label{fig:rho150_2}
\end{figure*}

Our results presented above can be affected by the adopted fitting procedure. 
To confirm the robustness of the results, 
we try to fit the subhalo's density profile with the cuspy profile expressed by
\begin{eqnarray}
\rho_{\rm{cuspy}}(r) &=& \frac{\rho_{\rm s}r_{\rm s}^{3}}{r^{\alpha}(r_{\rm s}+r)^{3-\alpha}}. 
\label{eq:cuspyprf}
\end{eqnarray}
This cuspy profile is the same as the NFW when $\alpha=1$.  The
central slope $\alpha$ is a fitting parameter~(with the range of $0 < \alpha < 3$), and thus, this profile
can be applied for both cuspy and cored profiles.  Figure
\ref{fig:rho150_2} shows the results using the cuspy profile for all
three simulations.  
The fitting procedure has little impact on the density \rhoc\ in the CDM case.
On the other hand, the density \rhoc\ in the SIDM simulations is systematically
higher in the cuspy profile than in the Burkert profile because
subhalos can have cores or very shallow cusps (shallower than $-1$).
However, the overall trend between \rhoc\ and \rperi\ in each
simulation is the same as Fig.~\ref{fig:rho150_1} and \ref{fig:rho150_2}.

\subsection{Density profile of subhalos}\label{sec_mass_grad}

As discussed in the previous section, the central density of subhalos
with small pericenter shows a stark difference between the
CDM and SIDM simulations. 
To understand the origin of the difference, 
we separate subhalos according to the pericenter 
and compare the density profile of each simulation.

\begin{figure*}
\includegraphics[width=15.0cm]{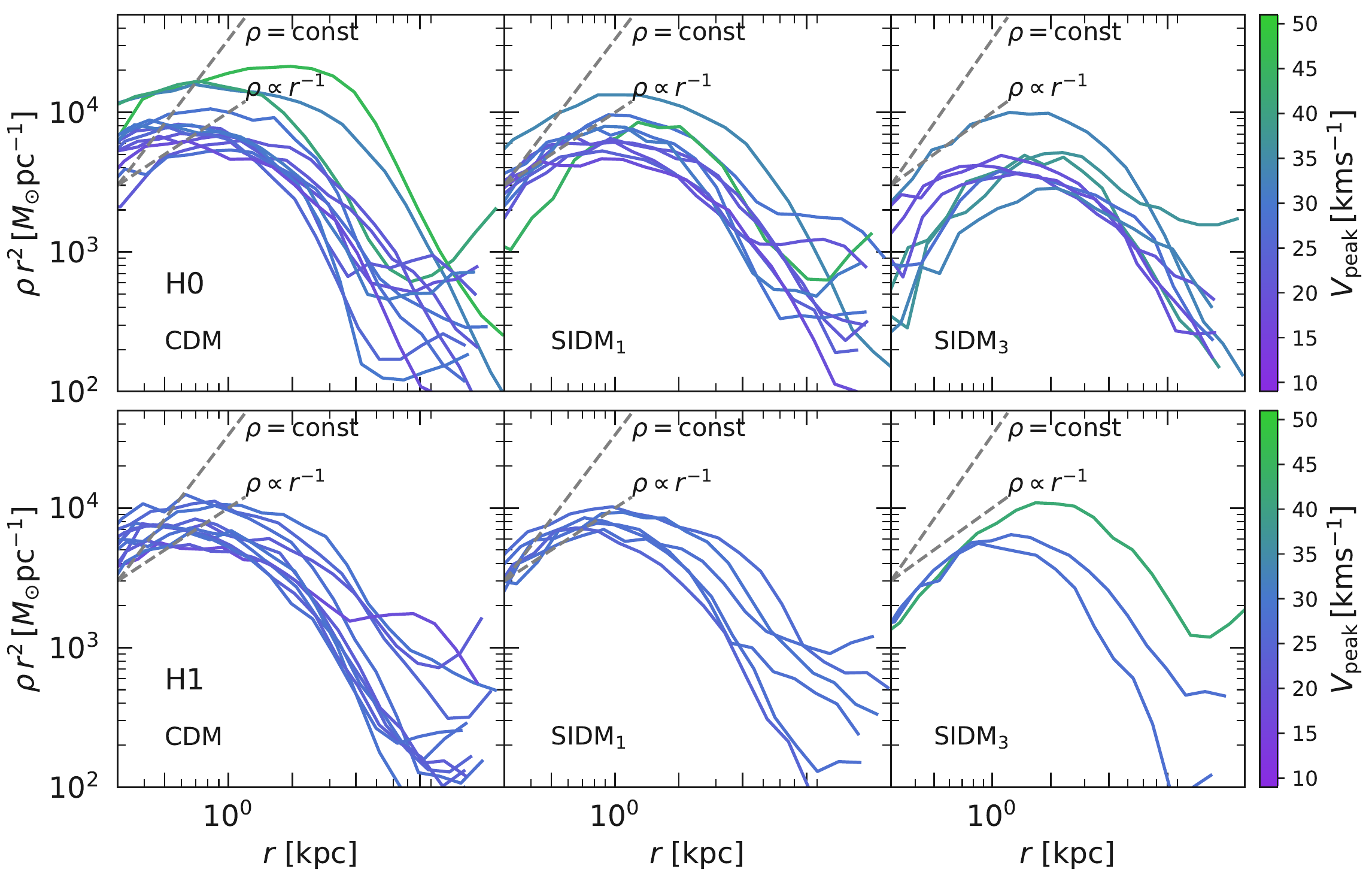}
\caption{
Density profiles of subhalos with $r_{\rm{peri}}<30\ \rm{kpc}$ 
for CDM (left), \sidmi\ (middle), and \sidmii\ (right). 
Each curve is colored in proportion to the \vpeak.
Top and bottom panels show the results of subhalos
hosted by H0 and H1, respectively.
}
\label{fig:mass_grad_1}
\end{figure*}

\begin{figure*}
\includegraphics[width=15.0cm]{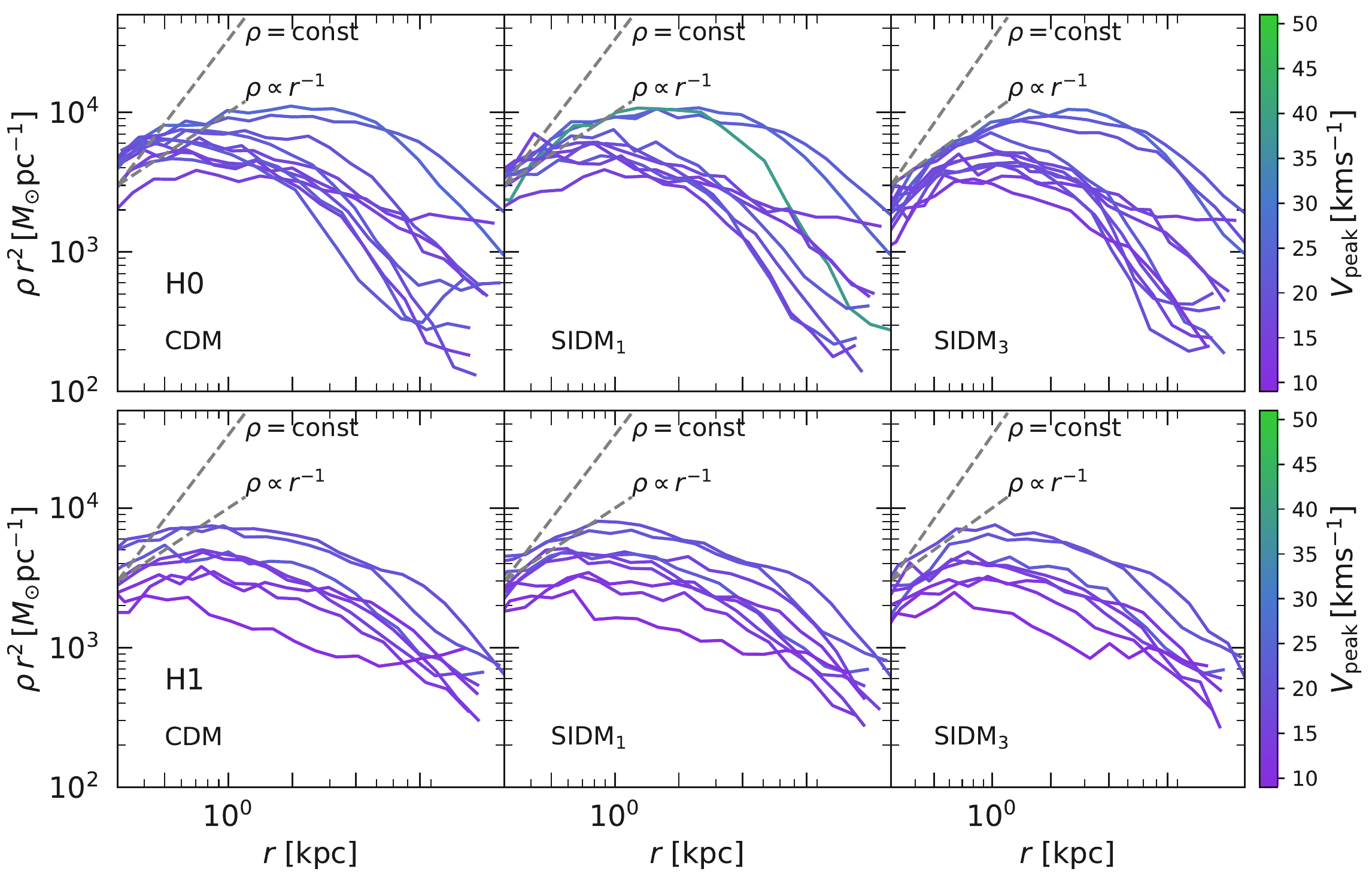}
\caption{
Same as Fig.~\ref{fig:mass_grad_1} but for subhalos with $130 \leq
r_{\rm{peri}} < 150\ \rm{kpc}$.
}
\label{fig:mass_grad_2}
\end{figure*}

Fig.~\ref{fig:mass_grad_1} and~\ref{fig:mass_grad_2} show density
profiles of subhalos with the pericenter $r_{\rm{peri}} < 30 \rm{kpc}$
and $130\ \rm{kpc} \leq r_{\rm{peri}} < 150\ \rm{kpc}$, respectively, 
in which the color of each curve is proportional to its \vpeak. 
We plot only subhalos with masses between $10^8$ and $10^9$\hMsun.
The top and bottom panels show the results of subhalos
hosted by H0 and H1, respectively.  From left to right, the results for the CDM, \sidmi,
and \sidmii\ simulations are shown.   For subhalos 
with small pericenter ($r_{\rm{peri}} < 30 \rm{kpc}$), 
the density slopes of the very
inner regions are steepest in the CDM than in the SIDM simulations.
The logarithmic slope $\frac{d\log \rho}{d\log r}$ is typically $-1$ 
for the CDM, between $-1$ and $0$
for the \sidmi, and $\sim 0$ (constant density)  
for the \sidmii, highlighting the effect of
self-interactions.  On the other hand, for subhalos far 
from the centers of host halos ($130\ \rm{kpc} \leq r_{\rm{peri}} <
150\ \rm{kpc}$), the difference between the CDM and SIDM simulations is
much smaller than for the inner regions. 
Furthermore, we also confirm these trends in other host halos (H2-8), 
supporting that those are not results of halo-to-halo variations. 

Cosmological CDM simulations suggest that subhalos with small
\rperi\ are subjected to stronger tidal forces from their hosts than
those with large \rperi, and thus, less dense subhalos (and also lower
\vpeak) with small \rperi\ could be 
destroyed~\cite{Kaplinghat2019, Hayashi2020}.  
Therefore, there are fewer subhalos with both small
\rperi\ and low \vpeak\ in the CDM simulation, which is clearly seen
in Fig.~\ref{fig:rho150_1} and~\ref{fig:rho150_2}, and by comparing
Fig.~\ref{fig:mass_grad_1} and~\ref{fig:mass_grad_2}.  On the other
hand, dark matter self-interactions between subhalo's particles
work effectively in the dense
central region of subhalos with higher \vpeak\ for the SIDM cases, and
hence, the central densities of such subhalos tend to be lower, and
their core tends to be larger.  
Dark matter self-interactions between subhalo's and host halo's particles
could further enhance tidal disruption of subhalos on radial 
orbits~\cite{Nadler2020}.
As a result, since the SIDM subhalos 
even with higher \vpeak\ could be destroyed by tidal effect, 
the dependence of \rhoc\ on the pericenter 
disappears in the SIDM simulations 
(see Fig.~\ref{fig:mass_grad_1} and~\ref{fig:mass_grad_2}), 
and the difference between the CDM and SIDM is prominent in subhalos
with small pericenter. 
These results highlight that the Milky-Way satellites with small pericenter are ideal
sites to further constrain the nature of dark matter by future large
spectroscopic surveys such as PFS~\cite{Takada2014} and Euclid
\cite{Euclid2011}.

\section{Discussion}\label{sec_discussion}

We discuss two caveats associated with our main results: gravothermal
core-collapse and galactic disk.  Gravothermal core-collapse can
increase the central density of subhalos although we do not observe
the sign in our simulations.  The Galactic disk potential, which is
not included in our simulations, could alter the evolution of subhalos
with small pericenter.

\subsection{Gravothermal core-collapse}
Dark matter self-interaction leads to an outer heat transfer, thereby
inducing gravothermal core-collapse such that the central density
increases with time~\citep{Balberg:2002ue,Nishikawa:2019lsc, Feng2021}.  
The core-collapse time scales of halos depend largely on
self-interaction cross-section and dynamical evolution of halos.
Provided that a halo is isolated and the SIDM cross-section is smaller
than $10\ \rm{cm^{2}g^{-1}}$, the core-collapse cannot occur within
the age of Universe.  However, subhalos we consider in this work have
experienced tidal stripping from the gravitational potential of a host
halo.  This tidal effect induces the core-collapse in the central
region of subhalos, and the time scale of the core-collapse can be
shorter than the age of the Universe, even though the cross-section is
smaller than $\sigma/m \sim
3\ \rm{cm^{2}g^{-1}}$~\citep{Kahlhoefer2019,Nishikawa:2019lsc,Sameie:2019zfo,Correa:2020qam,Kamada:2020buc}.
Nevertheless, there are no such dense subhalos even in the
\sidmii\ simulation~(Fig.~\ref{fig:rho150_1} and~\ref{fig:rho150_2}),
suggesting that $\sigma/m < 3\ \rm{cm^{2}g^{-1}}$ is insufficient to
induce core-collapse.  In the case of $\sigma/m >
3\ \rm{cm^{2}g^{-1}}$, this process can potentially mitigate the
discrepancy between SIDM models and the observed dSphs in
\rhoc~--~\rperi\ plane~\cite{Kaplinghat:2019svz,Kahlhoefer2019,Correa:2020qam}.
It, however, is to be noted that it is unclear whether the
core-collapse induced by tidal stripping occurs generally.  This is
because such core-collapse mechanism depends on halo concentration and
orbital properties (e.g., eccentricity, pericenter radius, infall
time, and so on).  Self-interaction between host halo and subhalo
particles can alter the abundance and dynamics of
subhalos~\cite{Nadler2020}.  Although its effect depends on SIDM
models and even how self-interaction is implemented in simulations, it
could also shorten the timescale of core-collapse.  Therefore, we need
further studies to assess the impact of core-collapse.

\subsection{Galactic disk}

The Galactic disk potential, which is not included in our simulations, could make the inner dark matter potential wall much deeper.
Thus, adding a stellar disk preferentially reduces the subhalo densities and the subhalo abundance 
with smaller pericenter distances (e.g.,~\cite{Kelley2019, Robles2019}), 
depending on the subhalo concentrations and orbital inclinations~\cite{Kahlhoefer2019}.
In particular, subhalos with pericenter less than 20$\sim$30 kpc is affected, 
therefore, the overall trend between \rhoc\ and \rperi\ in each of our simulations should not be altered.

\section{Conclusions}\label{sec_conclusions}
We conducted high-resolution cosmological $N$-body simulations based
on both CDM and SIDM that resolve density structures of the classical
dSph and massive UFD scale subhalos.  We have quantified the subhalos'
central density at 150 pc from the center of each subhalo.  Comparing
them with observational data, we have found that the SIDM scattering
cross-section of $\sigma/m<3\ \rm{cm^{2}g^{-1}}$ is favored. Subhalos
with higher \rhoc\ tend to have lower \vpeak\ in the \sidmi\ and
\sidmii\ simulations, whereas CDM counterparts tend to have higher
\vpeak.  This feature is prominent in subhalos with small pericenter due to the combination of hosts' tide and
self-interactions.  Therefore, the Milky-Way satellites with small pericenter are
ideal sites to further constrain the nature of dark matter by future
large spectroscopic surveys.

\section*{acknowledgments}
We thank Ethan O. Nadler for his helpful discussions.  Numerical
computations were carried out on Aterui-II supercomputer at the Center
for Computational Astrophysics, National Astronomical Observatory of
Japan.  This work has been supported by MEXT as ``Program for
Promoting Researches on the Supercomputer Fugaku'' (JPMXP1020200109),
and JICFuS.  We thank the support of MEXT/JSPS KAKENHI Grant Number
JP18H04337, JP20H05245 (for TI), 20H01895, and 21K13909 (for KH).


\appendix*
\section{Validation of the SIDM implementation}\label{sec:core}
\begin{figure}
\includegraphics[width=8cm]{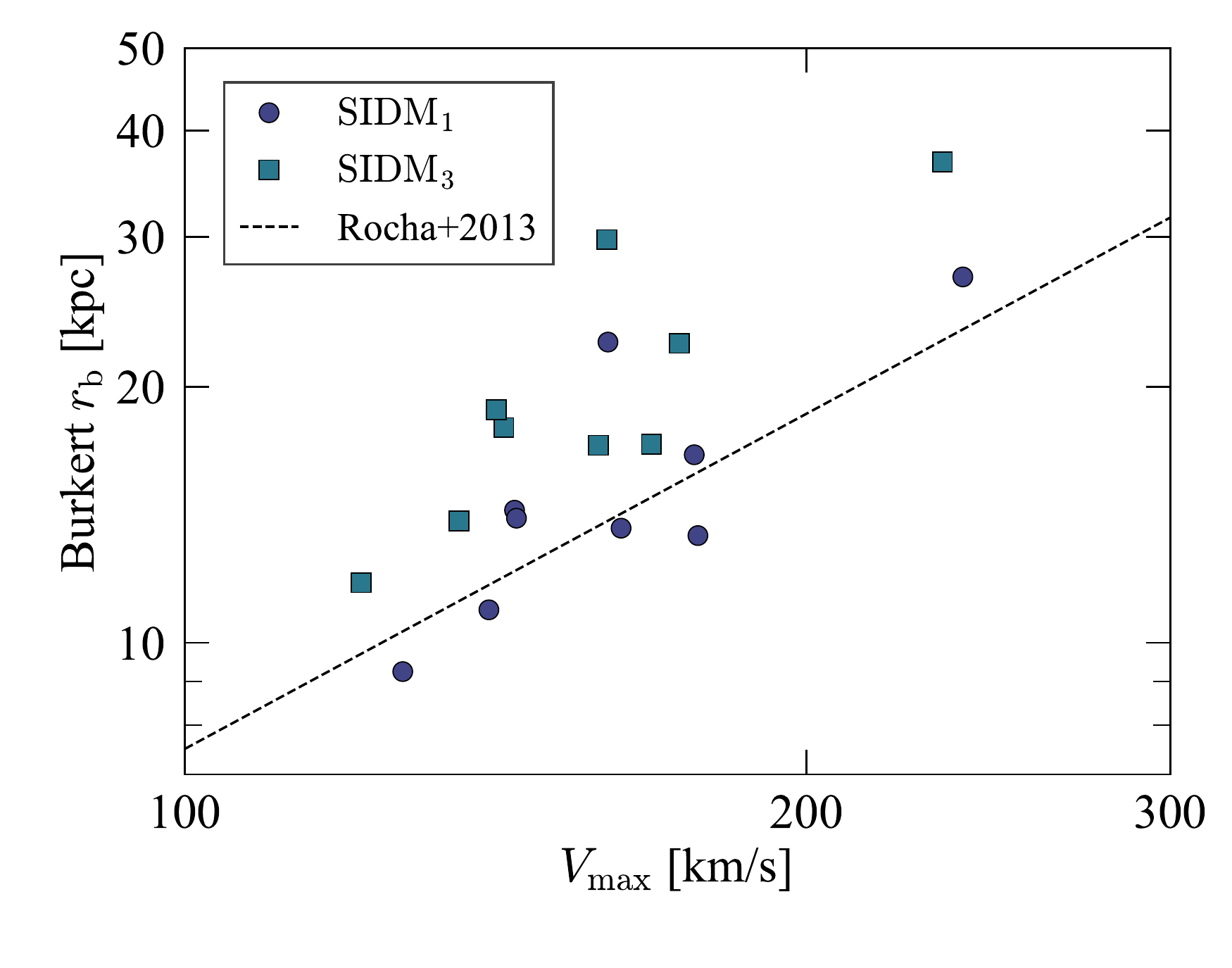}
\caption{
Burkert scale radius $r_{\rm b}$ 
of the nine Milky Way-sized host halos 
versus their circular velocity peak \vmax, 
for the \sidmi\ and \sidmii\ simulations. 
Dashed curve is a single power law fitting provided by 
\citet{Rocha2013} from their \sidmi\ simulations, 
$r_{\rm b} = 7.50 \, {\rm kpc} \left(\frac{V_{\rm max}}{\rm 100 km s^{-1}} \right)^{1.31}$
}
\label{fig:halo_core}
\end{figure}

In this section, we validate our SIDM implementation by comparing the
size of the SIDM core with a previous study.  We fitted the Burkert
profile to density profiles of the nine Milky Way-sized host halos
from the \sidmi\ and \sidmii\ simulations, and obtained scale radius
$r_{\rm b}$.  Fig.~\ref{fig:halo_core} shows $r_{\rm b}$ versus host
halos' circular velocity peak \vmax.  Our \sidmi\ halos are in good
agreement with a previous simulation study \citep{Rocha2013}, and the
core size is systematically larger in \sidmii\ halos than in
\sidmi\ halos, reinforcing the validity of our SIDM implementation.

\end{document}